\documentclass{article}
\usepackage{eurosym}
\usepackage{amssymb}
\usepackage{amsmath}
\usepackage{cite}
\usepackage{graphicx}
\usepackage{epsf}
\usepackage{lscape}

\begin{document}

\title{One-dimensional optimal system for 2D Rotating Ideal Gas}
\author{Andronikos Paliathanasis\thanks{%
Email: anpaliat@phys.uoa.gr} \\
{\ \ \textit{Institute of Systems Science, Durban University of Technology }}%
\\
{\ \textit{PO Box 1334, Durban 4000, Republic of South Africa}}}
\maketitle

\begin{abstract}
We derive the one-dimensional optimal system for a system of three partial
differential equations which describe the two-dimensional rotating ideal gas with polytropic parameter $\gamma >2.$ The Lie symmetries and the
one-dimensional optimal system are determined for the nonrotating and
rotating systems. We compare the results and we found that when there is no
Coriolis force the system admits eight Lie point symmetries, while the
rotating system admits seven Lie point symmetries. Consequently the two
systems are not algebraic equivalent as in the case of $\gamma =2~$ which was
found by previous studies. For the one-dimensional optimal system we
determine all the Lie invariants, while we demonstrate our results by
reducing the system of partial differential equations into a system of
first-order ordinary differential equations which can be solved by
quadratures.

\bigskip

\end{abstract}

\section{Introduction}

\label{sec1}

A powerful mathematical treatment for the determination of exact solutions
for nonlinear differential equations is the Lie symmetry analysis \cite%
{olver,kumei,ibra}. Specifically, Lie point symmetries help us in the
simplification of differential equations by means of similarity
transformations which reduce the differential equation. The reduction
process is based on the existence of functions which are invariant under a
specific group of point transformations. When someone uses these invariants
as new dependent and independent variables the differential equation is
reduced. The reduction process differs between ordinary differential
equations (ODEs) and partial differential equations (PDEs). For ODEs Lie
point\ symmetries are applied to reduce the order of ODE by one; while on
PDEs Lie point symmetries are applied to reduce by one the number of
independent variables, while the order of the PDE remain the same. The
solutions which are found with the application of those invariant functions
are called similarity solutions. Some applications on the determination of
similarity solutions for nonlinear differential equations can be found in
\cite{ref1,ref2,ref3,ref4,ref5,ref10} and references therein.

A common characteristic in the reduction process is that the Lie point
symmetries are not preserved during the reduction, hence we can say that the
symmetries can be lost. That is not an accurate statement, because
symmetries are not \textquotedblleft destroyed\textquotedblright\ or
\textquotedblleft created\textquotedblright\ under point transformations but
the \textquotedblleft nature\textquotedblright\ of the symmetry change. In
addition, Lie symmetries can be used to construct new similarity solutions
for a given differential equation by applying the adjoint representation of
the Lie group \cite{leach01}.

It is possible that a given differential equation to admit more than one
similarity solution when the given differential equation admits a
\textquotedblleft large\textquotedblright\ number of Lie point symmetries.
Hence, in order for someone to classify a differential equation according to
the admitted similarity solutions, all the inequivalent Lie subalgebras of
the admitted Lie symmetries should be determined.

The first group classification problem was carried out by Ovsiannikov \cite%
{Ovsi} who demonstrated the construction of the one-dimensional optimal
system for the Lie algebra. Since then, the classification of the
one-dimensional optimal system has become a main tool for the study of
nonlinear differential equations \cite{opt1,opt2,opt3,opt4}.

In this work, we focus on the classification of the one-dimensional optimal
system for the two-dimensional rotating ideal gas system for a described by
the following system of PDEs \cite{val1,cc1,kk1}

\begin{eqnarray}
h_{t}+\left( hu\right) _{x}+\left( hv\right) _{y} &=&0,  \label{sw.01} \\
u_{t}+uu_{x}+vu_{y}+h^{\gamma -2}h_{x}-fv &=&0,  \label{sw.02} \\
v_{t}+uv_{x}+vv_{y}+h^{\gamma -2}h_{y}+fu &=&0.  \label{sw.03}
\end{eqnarray}%
where $u$ and $v$ are the velocity components on the $x$ and $y$ directions
respectively, $h$ is the density of the ideal gas, $f$ is the Coriolis
parameter and $\gamma $ is the polytropic parameter of the fluid. Usually $%
\gamma $ is assumed to be $\gamma =2$ where equations (\ref{sw.01})-(\ref%
{sw.03}) reduce to the shallow water system. However in this work we
consider that $\gamma >2$. In this work, polytropic index $\gamma $ is
defined as $\frac{C_{p}}{C_{v}}=\gamma -1$.

Shallow-water equations describe the flow of a fluid under a pressure
surface. There are various physical phenomena which are described by the
Shallow-water system with emphasis on atmospheric and oceanic phenomena \cite%
{oc01,oc02,oc03}. Hence, the existence of the Coriolis force it becomes
critical in the description of the physical phenomena.

In the case of $\gamma =2$, the complete symmetry analysis of the system (%
\ref{sw.01})-(\ref{sw.03}) is presented in \cite{swr06}. It was found that
for $\gamma =2$ the given system of PDEs is invariant under a
nine-dimensional Lie algebra. The same Lie algebra but in a different
representation is also admitted by the nonrotating system, i.e. $f=0$.\ One
of the main results of \cite{swr06} is that the transformation which relates
the two representations of the admitted Lie algebras for the rotating and
nonrotating system, transform the rotating system (\ref{sw.01})-(\ref{sw.03}%
) into the nonrotating one. For other applications of Lie symmetries on
Shallow-water equation we refer the reader in \cite{sw1,sw2,sw3,sw4,sw5,sw6}.

For the case of a ideal gas \cite{cc1}, i.e. parameter $\gamma >1$ from our
analysis it follows that this property is lost. The nonrotating system and
the rotating one are invariant under a different number of Lie symmetries
and consequently under different Lie algebras. For each of the Lie algebras
we \ae ther the one-dimensional optimal system and all the Lie invariants.
The results are presented in tables. We demonstrate the application of the
Lie invariants by determining some similarity solutions for the system (\ref%
{sw.01})-(\ref{sw.03}) for $\gamma >2$. \ The paper is structured as follows.

In Section \ref{sec2}. we briefly discuss the theory of Lie symmetries for
differential equations and the adjoint representation. The nonrotating
system (\ref{sw.01})-(\ref{sw.03}) is studied in Section \ref{sec3}.
Specifically we determine the Lie points symmetries which form an
eight-dimensional Lie algebra. The commutators and the adjoint
representation are presented. We make use of these results and we perform a
classification of the one-dimensional optimal system. We found that in total
there are twenty-three one-dimensional indepedent Lie symmetries and
possible reductions, the corresponding invariants are determined and
presented in tables. In Section \ref{sec4} we perform the same analysis for
the rotating system. There we find that the admitted Lie symmetries form a
seven-dimensional Lie algebra while there are twenty independent
one-dimensional Lie algebras. We demonstrate the results by reducing the
system of PDEs (\ref{sw.01})-(\ref{sw.03}) into an integrable system of
three first-order ODEs which solution is given by quadratures. In Section %
\ref{sec5} we discuss our results and draw our conclusions. Finally, in
Appendix \ref{app1} we present the tables which includes the results of our
analysis.

\section{Lie symmetry analysis}

\label{sec2}

Let $H^{A}\left( x^{i},\Phi ^{A},\Phi _{i}^{A},...\right) =0$ be a system of
partial differential equations (PDEs) where $\Phi ^{A}~$denotes the
dependent variables and $x^{i}$ are the indepedent variables, at this point
it is important to mention that we make use of the Einstein summation
convention. By definition under the action of the infinitesimal
one-parameter point transformation (1PPT)
\begin{equation}
\bar{x}^{i}=x^{i}\left( x^{j},\Phi ^{B};\varepsilon \right) ,~~\bar{\Phi}%
^{A}=\Phi ^{A}\left( x^{j},\Phi ^{B};\varepsilon \right) ,  \label{de.03}
\end{equation}%
which connects two different points $P\left( x^{j},\Phi ^{B}\right)
\rightarrow Q\left( \bar{x}^{j},\bar{\Phi}^{B},\varepsilon \right) $, the
differential equation $H^{A}=0$ remains invariant if and only if $\bar{H}%
^{A}=H^{A}$, that is \cite{kumei}%
\begin{equation}
\lim_{\varepsilon \rightarrow 0}\frac{\bar{H}^{A}\left( \bar{y}^{i},\bar{u}%
^{A},...;\varepsilon \right) -H^{A}\left( y^{i},u^{A},...\right) }{%
\varepsilon }=0.  \label{ls.05}
\end{equation}%
The latter condition means that the $\Phi ^{A}\left( P\right) $ and $\Phi
^{A}\left( Q\right) $ are connected through the transformation.

The lhs of expression (\ref{ls.05}) defines the Lie derivative of $H^{A}$
along the vector field $X$ of the one-parameter point transformation (\ref%
{de.03}), in which $X$ is defined as
\begin{equation*}
X=\frac{\partial \bar{x}^{i}}{\partial \varepsilon }\partial _{i}+\frac{%
\partial \bar{\Phi}}{\partial \varepsilon }\partial _{A}.
\end{equation*}

Thus, condition (\ref{ls.05}) is equivalent with the following expression
\cite{kumei}%
\begin{equation}
\mathcal{L}_{X}\left( H^{A}\right) =0,  \label{ls.05a}
\end{equation}%
where~$\mathcal{L}$ denotes the Lie derivative with respect to the vector
field $X^{\left[ n\right] }$ which is the $n$th-extension of generator $X~$%
of the transformation (\ref{de.03}) in the jet space $\left\{ x^{i},\Phi
^{A},\Phi _{,i}^{A},\Phi _{,ij}^{A},...\right\} $ given by the expression
\cite{kumei}
\begin{equation}
X^{\left[ n\right] }=X+\eta ^{\left[ 1\right] }\partial _{\Phi
_{i}^{A}}+...+\eta ^{\left[ n\right] }\partial _{\Phi
_{i_{i}i_{j}...i_{n}}^{A}},  \label{ls.06}
\end{equation}%
in which%
\begin{equation}
\eta ^{\left[ n\right] }=D_{i}\eta ^{\left[ n-1\right]
}-u_{i_{1}i_{2}...i_{n-1}}D_{i}\left( \frac{\partial \bar{x}^{j}}{\partial
\varepsilon }\right) ~,~i\succeq 1~,~\eta ^{\left[ 0\right] }=\left( \frac{%
\partial \bar{\Phi}^{A}}{\partial \varepsilon }\right) .  \label{de.08}
\end{equation}

Conditions (\ref{ls.05a}) provides a system of PDEs whose solution determine
the components of the $X$, consequently the infinitesimal transformation.
The vector fields $X$ which satisfy condition (\ref{ls.05a}) are called Lie
symmetries for the differential equation $H^{A}=0$. The Lie symmetries for a
given differential equation form a Lie algebra.

Lie symmetries can be used by different ways\ \cite{kumei} in order to study
a differential equation. However, their direct application is on the
determination of the so-called similarity solutions. The steps which we
follow to determine a similarity solution is based on the determination and
application of the Lie invariant functions.

Let $X$ be a Lie symmetry for a given differential equation~$H^{A}=0$, then
the differential equation $X\left( F\right) =0$, where $F$ is a function,
provides the Lie invariants where by replacing in the differential equation $%
H^{A}=0$, we reduce the number of the indepedent variables (in the case of
PDEs) or the order of the differential equation (in the case of ordinary
differential equations (ODEs)).

\subsection{Optimal system}

Consider the $n$-dimensional Lie algebra $G_{n}$ with elements $%
X_{1},~X_{2},~...~X_{n}$. Then we shall say that the two vector fields \cite%
{kumei}
\begin{equation}
Z=\sum\limits_{i=1}^{n}a_{i}X_{i}~,~W=\sum\limits_{i=1}^{n}b_{i}X_{i}~,~%
\text{\ }a_{i},~b_{i}\text{ are constants.}  \label{sw.04}
\end{equation}%
are equivalent iff there
\begin{equation}
\mathbf{W}=lim_{j=i}^{n}Ad\left( \exp \left( \varepsilon _{i}X_{i}\right)
\right) \mathbf{Z}  \label{sw.05}
\end{equation}%
or%
\begin{equation}
W=cZ~,~c=const.  \label{sw.06}
\end{equation}%
where the operator \cite{kumei}%
\begin{equation}
Ad\left( \exp \left( \varepsilon X_{i}\right) \right)
X_{j}=X_{j}-\varepsilon \left[ X_{i},X_{j}\right] +\frac{1}{2}\varepsilon
^{2}\left[ X_{i},\left[ X_{i},X_{j}\right] \right] +...  \label{sw.07}
\end{equation}%
is called the adjoint representation.$~$

Therefore, in order to perform a complete classification for the similarity
solutions of a given differential equation we should determine all the
one-dimensional indepedent symmetry vectors of the Lie algebra $G_{n}$.

We continue our analysis by calculating the Lie point symmetries for the
system (\ref{sw.01})-(\ref{sw.03}) for the case where the system is rotating$%
~(f\neq 0)\ $and nonrotating $\left( f=0\right) $.

\section{Symmetries and optimal system for nonrotating shallow water}

\label{sec3}

We start our analysis by applying the symmetry condition (\ref{ls.05a}) for
the Coriolis free system (\ref{sw.01})-(\ref{sw.03}) with $f=0$. We found
that the system of PDEs admit eight Lie point symmetries as they are
presented in the following \cite{Ovsi}
\begin{eqnarray*}
X_{1} &=&\partial _{t}~,~X_{2}=\partial _{x}~,~X_{3}=\partial _{y}~, \\
X_{4} &=&t\partial _{x}+\partial _{u}~,~X_{5}=t\partial _{y}+\partial _{v}~,
\\
X_{6} &=&y\partial _{x}-x\partial _{y}+v\partial _{u}-u\partial _{v}, \\
X_{7} &=&t\partial _{t}+x\partial _{x}+y\partial _{y}, \\
X_{8} &=&\left( \gamma -1\right) \left( x\partial _{x}+y\partial
_{y}+u\partial _{u}+v\partial _{v}\right) +2h\partial _{h}.
\end{eqnarray*}

The commutators of the Lie symmetries and the adjoint representation are
presented in tables \ref{tabl1} and \ref{tabl2} respectively.

\begin{table}[tbp] \centering%
\caption{Commutators of the admitted Lie point symmetries for the
nonrotating 2D Shallow water}%
\begin{tabular}{ccccccccc}
\hline\hline
$\left[ ~,~\right] $ & $\mathbf{X}_{1}$ & $\mathbf{X}_{2}$ & $\mathbf{X}_{3}$
& $\mathbf{X}_{4}$ & $\mathbf{X}_{5}$ & $\mathbf{X}_{6}$ & $\mathbf{X}_{7}$
& $\mathbf{X}_{8}$ \\
$\mathbf{X}_{1}$ & $0$ & $0$ & $0$ & $X_{2}$ & $X_{3}$ & $0$ & $-\left(
\gamma -1\right) X_{1}$ & $0$ \\
$\mathbf{X}_{2}$ & $0$ & $0$ & $0$ & $0$ & $0$ & $-X_{3}$ & $0$ & $\left(
\gamma -1\right) X_{2}$ \\
$\mathbf{X}_{3}$ & $0$ & $0$ & $0$ & $0$ & $0$ & $X_{2}$ & $0$ & $\left(
\gamma -1\right) X_{3}$ \\
$\mathbf{X}_{4}$ & $-X_{2}$ & $0$ & $0$ & $0$ & $0$ & $-X_{5}$ & $\left(
\gamma -1\right) X_{4}$ & $\left( \gamma -1\right) X_{4}$ \\
$\mathbf{X}_{5}$ & $-X_{3}$ & $0$ & $0$ & $0$ & $0$ & $X_{4}$ & $\left(
\gamma -1\right) X_{5}$ & $\left( \gamma -1\right) X_{5}$ \\
$\mathbf{X}_{6}$ & $0$ & $X_{3}$ & $-X_{2}$ & $X_{5}$ & $-X_{4}$ & $0$ & $0$
& $0$ \\
$\mathbf{X}_{7}$ & $\left( \gamma -1\right) X_{1}$ & $0$ & $0$ & $-\left(
\gamma -1\right) X_{4}$ & $-\left( \gamma -1\right) X_{5}$ & $0$ & $0$ & $0$
\\
$\mathbf{X}_{8}$ & $0$ & $-\left( \gamma -1\right) X_{2}$ & $-\left( \gamma
-1\right) X_{3}$ & $-\left( \gamma -1\right) X_{4}$ & $-\left( \gamma
-1\right) X_{5}$ & $0$ & $0$ & $0$ \\ \hline\hline
\end{tabular}%
\label{tabl1}%
\end{table}%

We continue by determine the one-dimensional optimal system. Let us consider
the generic symmetry vector%
\begin{equation*}
Z^{8}=a_{1}X_{1}+a_{2}X_{2}+a_{3}X_{3}+a_{4}X_{4}+a_{5}X_{5}+a_{6}X_{6}+a_{7}X_{7}+a_{8}X_{8}
\end{equation*}%
From table \ref{tabl2} we see that by applying the following adjoint
representations%
\begin{equation*}
Z^{\prime 8}=Ad\left( \exp \left( \varepsilon _{5}X_{5}\right) \right)
Ad\left( \exp \left( \varepsilon _{4}X_{4}\right) \right) Ad\left( \exp
\left( \varepsilon _{3}X_{3}\right) \right) Ad\left( \exp \left( \varepsilon
_{2}X_{2}\right) \right) Ad\left( \exp \left( \varepsilon _{1}X_{1}\right)
\right) Z^{8}
\end{equation*}%
parameters $\varepsilon _{1},~\varepsilon _{2},~\varepsilon _{3}$, $%
\varepsilon _{4}$ and~$\varepsilon _{5}$ can be determined such that%
\begin{equation*}
Z^{\prime 8}=a_{6}^{\prime }X_{6}+a_{7}^{\prime }X_{7}+a_{8}^{\prime }X_{8}
\end{equation*}

Parameters $a_{6},~a_{7},$ and $a_{8}$ are the relative invariants of the
full adjoint action. Indeed in order to determine the relative invariants we
solve the following system of partial differential equation \cite{olver}%
\begin{equation*}
\Delta \left( \phi \left( a_{i}\right) \right) =C_{ij}^{k}a^{i}\frac{%
\partial }{\partial a_{j}}
\end{equation*}%
where $C_{ij}^{k}$ are the structure constants of the admitted Lie algebra
as they presented in Table \ref{tabl1}. Consequently, in order to derive all
the possible one-dimensional Lie symmetries we should study various cases
were non of the invariants are zero, one of the invariants are zero, two of
the invariants are zero or all the invariants are zero.

Hence, for the first tree cases infer the following one-dimensional
independent Lie algebras%
\begin{equation*}
X_{6}~,~X_{7}~,~X_{8}~,~\xi _{\left( 67\right) }=X_{6}+\alpha X_{7}~,~\xi
_{\left( 68\right) }=X_{6}+\alpha X_{8}~
\end{equation*}%
\begin{equation*}
\xi _{\left( 78\right) }=X_{7}+\alpha X_{8}~,~\xi _{\left( 678\right)
}=X_{6}+\alpha X_{7}+\beta X_{8}.
\end{equation*}

We apply the same procedure for the rest of the possible linear combinations
of the symmetry vectors and we find the one-dimensional dependent Lie
algebras%
\begin{equation*}
X_{1},~X_{2}~,~X_{3}~,~X_{4}~,~X_{5}~,~\xi _{\left( 12\right) }=X_{1}+\alpha
X_{2}~,~\xi _{\left( 13\right) }=X_{1}+\alpha X_{3}~,~\xi _{\left( 23\right)
}=X_{2}+\alpha X_{3}~,~\xi _{\left( 14\right) }=X_{1}+\alpha X_{4}~,
\end{equation*}%
\begin{equation*}
~\xi _{\left( 15\right) }=X_{1}+\alpha X_{5}~,~\xi _{\left( 16\right)
}=X_{1}+\alpha X_{6},~\xi _{\left( 34\right) }=X_{3}+\alpha X_{4}~,~\xi
_{\left( 25\right) }=X_{2}+\alpha X_{5}~~\xi _{\left( 45\right)
}=X_{4}+\alpha X_{5}~,
\end{equation*}%
\begin{equation*}
~\xi _{\left( 123\right) }=X_{1}+\alpha X_{2}+\beta X_{3}~\xi _{\left(
145\right) }=X_{1}+\alpha X_{4}+\beta X_{5}~,~\xi _{\left( 125\right)
}=X_{1}+\alpha X_{2}+\beta X_{5}~,~\xi _{\left( 134\right) }=X_{1}+\alpha
X_{3}+\beta X_{4},
\end{equation*}%
in which $\alpha $ and $\beta $ are constants.

Therefore by applying one of the above Lie symmetry vectors we find all the
possible reductions from a system of $1+2$ PDEs to a system of $1+1$ PDEs.
The reduced system will not admit all the remaining Lie symmetries. The Lie
symmetries which survive under a reduction process are given as described in
the following example.

Let a PDE admits the Lie point symmetries $\Gamma _{1},~\Gamma _{2}$ which
are such that $\left[ \Gamma _{1},\Gamma _{2}\right] =C_{12}^{1}X_{1},~$with
$C_{12}^{1}\neq 0$.$~$Reduction with the symmetry vector $\Gamma _{1}$ leads
to a reduced differential equation which admits $\Gamma _{2}$ as Lie
symmetry. On the other hand, reduction of the mother equation with respect
to the Lie symmetry $\Gamma _{2}$ leads to a different reduced differential
equation which does not admit as a Lie point symmetry the vector field $%
\Gamma _{1}.$ In case the two Lie symmetries form an Abelian Lie algebra,
i.e.~$C_{12}^{1}=0$, then under any reduction process symmetries are
preserved by any reduction.

We found that the optimal system admit twenty-three one-dimensional Lie
symmetries, and possible independent reductions. All the possible
twenty-three Lie invariants are presented in tables \ref{tabl2a} and \ref%
{tabl2b}.

An application of the Lie invariants is presented below.

\subsection{Application of $\protect\xi _{145}$}

Let us now demonstrate the results of tables \ref{tabl2a} and \ref{tabl2b}
by the Lie invariants of the symmetry vector $\xi _{145}$ and construct the
similarity solution for the system.

The application of $\xi _{145}$ in the nonrotating system (\ref{sw.01})-(\ref%
{sw.03}) reduce the PDEs in the following system
\begin{eqnarray}
\left( hu\right) _{z}+\left( hv\right) _{w} &=&0  \label{s.01} \\
\alpha +uu_{z}+vu_{w}+h^{\gamma -2}h_{z} &=&0  \label{s.02} \\
\beta +uv_{z}+vv_{w}+h^{\gamma -2}h_{w} &=&0  \label{s.03}
\end{eqnarray}%
where $z=x-\frac{\alpha }{2}t^{2}~$and $w=~y-\frac{\beta }{2}t^{2}$.

System (\ref{s.01})-(\ref{s.03}) admits the Lie point symmetries%
\begin{equation}
\partial _{z}~,~\partial _{w}~,~z\partial _{z}+w\partial _{w}+\frac{2}{%
\gamma -1}h\partial _{h}+u\partial _{u}+v\partial _{v}  \label{s.04}
\end{equation}

Reduction with the symmetry vector $\partial _{z}+c\partial _{w}$ provides
the following system of first-order ODEs%
\begin{eqnarray}
Fh_{\sigma } &=&\left( c\alpha -\beta \right) h^{2},  \label{s.05} \\
Fv_{\sigma } &=&\frac{\left( \alpha -c\beta \right) ch^{\gamma }-\alpha
h\left( v-cu\right) ^{2}}{v-cu},  \label{s.06} \\
Fh_{\sigma } &=&\frac{\left( \alpha -c\beta \right) cu^{\gamma }-\beta
h\left( v-cu\right) ^{2}}{v-cu}.  \label{s.08}
\end{eqnarray}%
where $F=\left( 1+c^{2}\right) h^{\gamma }-h\left( v-cu\right) ^{2}~$and $%
\sigma =z+cw$.

By performing the change of variable $d\sigma =fd\tau $ function $f$ can be
removed from the above system. For $h\left( \tau \right) =0$, system (\ref%
{s.05}), (\ref{s.06}), (\ref{s.08}) admits a solution $u=u_{0},~v=v_{0}$
which is a critical point. The latter special solutions are always unstable
when $\alpha c>\beta .$

We proceed our analysis by considering the rotating system.

\begin{table}[tbp] \centering%
\caption{Commutators of the admitted Lie point symmetries for the rotating
2D Shallow water}%
\begin{tabular}{cccccccc}
\hline\hline
$\left[ ~,~\right] $ & $\mathbf{Y}_{1}$ & $\mathbf{Y}_{2}$ & $\mathbf{Y}_{3}$
& $\mathbf{Y}_{4}$ & $\mathbf{Y}_{5}$ & $\mathbf{Y}_{6}$ & $\mathbf{Y}_{7}$
\\
$\mathbf{Y}_{1}$ & $0$ & $0$ & $0$ & $0$ & $fY_{6}$ & $-fY_{5}$ & $0$ \\
$\mathbf{Y}_{2}$ & $0$ & $0$ & $0$ & $-Y_{3}$ & $0$ & $0$ & $\left( \gamma
-1\right) Y_{2}$ \\
$\mathbf{Y}_{3}$ & $0$ & $0$ & $0$ & $Y_{2}$ & $0$ & $0$ & $\left( \gamma
-1\right) Y_{3}$ \\
$\mathbf{Y}_{4}$ & $0$ & $Y_{3}$ & $-Y_{2}$ & $0$ & $-Y_{6}$ & $Y_{5}$ & $0$
\\
$\mathbf{Y}_{5}$ & $-fY_{6}$ & $0$ & $0$ & $Y_{6}$ & $0$ & $0$ & $\left(
\gamma -1\right) Y_{5}$ \\
$\mathbf{Y}_{6}$ & $fY_{5}$ & $0$ & $0$ & $-Y_{5}$ & $0$ & $0$ & $\left(
\gamma -1\right) Y_{6}$ \\
$\mathbf{Y}_{7}$ & $0$ & $-\left( \gamma -1\right) Y_{2}$ & $-\left( \gamma
-1\right) Y_{3}$ & $0$ & $-\left( \gamma -1\right) Y_{5}$ & $-\left( \gamma
-1\right) Y_{6}$ & $0$ \\ \hline\hline
\end{tabular}%
\label{tabl3}%
\end{table}%

\section{Symmetries and optimal system for rotating shallow water}

\label{sec4}

For the rotating system $\left( f\neq 0\right) $, the Lie symmetries are
\begin{eqnarray*}
Y_{1} &=&\partial _{t}~,~Y_{2}=\partial _{x}~,~Y_{3}=\partial _{y}~, \\
Y_{4} &=&y\partial _{x}-x\partial _{y}+v\partial _{u}-u\partial _{v}~, \\
Y_{5} &=&\sin \left( ft\right) \partial _{x}+\cos \left( ft\right) \partial
_{y}+f\left( \cos \left( ft\right) \partial _{u}-\sin \left( ft\right)
\partial _{v}\right) \\
Y_{6} &=&\cos \left( ft\right) \partial _{x}-\sin \left( ft\right) \partial
_{y}-f\left( \sin \left( ft\right) \partial _{u}+\cos \left( ft\right)
\partial _{v}\right) \\
Y_{7} &=&\left( \gamma -1\right) \left( x\partial _{x}+y\partial
_{y}+u\partial _{u}+v\partial _{v}\right) +2h\partial _{h}
\end{eqnarray*}

The commutators and the adjoint representation are given in tables \ref%
{tabl3} and \ref{tabl4}. The Lie symmetries for the rotating system form a
smaller dimension Lie algebra than the non-rotating system. That is not the
case when $\gamma =2$, where the two Lie algebras have the same dimensional
and are equivalent under point transformation \cite{swr06}. Therefore, for $%
\gamma >2$ the Coriolis force cannot be eliminated by a point transformation
as in the $\gamma =2$ case.

As far as the admitted Lie symmetries admitted by the given system of PDEs
with or without the Coriolis terms for $\gamma >2$, we remark that the
rotating and the nonrotating system have a common Lie subalgebra of
one-parameter point transformations consists by the symmetry vectors $%
Y_{1},~Y_{2},~Y_{3},~Y_{4}$ and $Y_{7}$, or for the nonrotating system $%
X_{1},~X_{2},~X_{3},~X_{6}$ and $X_{8}$.

We proceed with the determination of the one-dimensional optimal system and
the invariant functions. Specifically, the relative invariants for the
adjoint representation are calculated to be $a_{1}~,~a_{7}$ and $a_{8}.~$%
From tables \ref{tabl3} and \ref{tabl4} we can find the one-dimensional
optimal system, which is
\begin{equation*}
Y_{1},~Y_{2},~Y_{3},~Y_{4},~Y_{5},~Y_{6},~Y_{7},~\chi _{12}=Y_{1}+\alpha
Y_{2},~\chi _{13}=Y_{1}+\alpha Y_{3},
\end{equation*}%
\begin{equation*}
\chi _{14}=Y_{1}+\alpha Y_{4}~,~\chi _{15}=Y_{1}+\alpha Y_{5},~~\chi
_{16}=Y_{1}+\alpha Y_{6},~\chi _{17}=Y_{1}+\alpha Y_{7},
\end{equation*}%
\begin{equation*}
\chi _{23}=Y_{2}+\alpha Y_{3}~,~~\chi _{45}=Y_{4}+\alpha Y_{5},~\chi
_{46}=Y_{4}+\alpha Y_{6},~\chi _{56}=Y_{5}+\alpha Y_{6}
\end{equation*}%
\begin{equation*}
~\chi _{47}=Y_{4}+\alpha Y_{6}~,~\chi _{123}=Y_{1}+\alpha Y_{2}+\beta
Y_{3},~\chi _{147}=Y_{1}+\alpha Y_{4}+\beta Y_{7}.
\end{equation*}

The Lie invariants which correspond to all the above one-dimensional Lie
algebras are presented in tables \ref{tabl4a} and \ref{tabl4b}.

Let us demonstrate the application of the Lie invariants by the following
applications, from where we can see that the Lie invariants reduce the
nonlinear field equations into a system of integrable first-order ODE which
can be solved with quadratures.

\subsection{Application of $\protect\chi _{12}$}

We consider the travel-wave similarity solution in the $x-$plane provided by
the symmetry vector $\chi _{12}$, and the vector field $Y_{3}$. The
resulting equations is described by the following system of first order ODEs%
\begin{eqnarray}
v_{z} &=&f\frac{u}{\alpha -u}~  \label{se.01} \\
\bar{F}u_{z} &=&f\left( \alpha -u\right) vh  \label{se.02} \\
\bar{F}h_{z} &=&fvh^{2}  \label{se.04}
\end{eqnarray}%
where $\bar{F}=h^{\gamma }-\left( a-u\right) ^{2}h$ and $z=t-\alpha x$.
Because we performed reduction with a subalgebra admitted by the nonrotating
system, by setting $f=0$ in (\ref{se.01}), (\ref{se.02}), (\ref{se.04}) we
get the similarity solution for the nonrotating system where in this case is
found to be $h\left( z\right) =h_{0},~u\left( z\right) =u_{0}$ and $v\left(
z\right) =v_{0}.$

We perform the substitution $dz=\frac{\bar{F}}{fv}d\tau $ and the latter
system is simplified as follows%
\begin{eqnarray}
\frac{v}{\bar{F}}v_{\tau } &=&\frac{u}{\alpha -u}~ \\
u_{\tau } &=&\left( \alpha -u\right) h \\
h_{\tau } &=&h^{2}
\end{eqnarray}%
from where we get the solution%
\begin{equation}
h\left( \tau \right) =\left( h_{0}-\tau \right) ^{-1}~,~u\left( \tau \right)
=\alpha +u_{0}-\frac{u_{0}}{h_{0}}\tau
\end{equation}%
and%
\begin{equation}
v\left( t\right) ^{2}=2\int \frac{\left( a+u_{0}-\frac{u_{0}}{h_{0}}\tau
\right) }{\frac{u_{0}}{h_{0}}\left( h_{0}-\tau \right) }\left( \left(
h_{0}-\tau \right) ^{-\gamma }+\left( \frac{u_{0}}{h_{0}}\right) ^{2}\tau -%
\frac{\left( u_{0}\right) ^{2}}{h_{0}}\right) d\tau .
\end{equation}

\subsection{Application of $\protect\chi _{23}$}

Consider now the reduction with the symmetry vector fields $\chi _{23}$. The
resulting system of 1+1 differential equations admit five Lie point
symmetries, they are%
\begin{eqnarray*}
&&\partial _{t},~\partial _{w}~,~\left( \sin \left( ft\right) +\alpha \cos
\left( ft\right) \right) \partial _{w}+f\left( \sin \left( ft\right)
\partial _{u}+\cos \left( ft\right) \partial _{v}\right) \\
&&\left( \alpha \sin \left( ft\right) -\cos \left( ft\right) \right)
\partial _{w}-f\left( \cos \left( ft\right) \partial _{u}-\sin \left(
ft\right) \partial _{v}\right) ~,~\left( \gamma -1\right) \left( \partial
_{w}+u\partial _{u}+v\partial _{v}\right) +2h\partial _{h}.
\end{eqnarray*}%
where $w=y-\alpha x$. For simplicity on our calculations let us assume $%
\gamma =3$.

Reduction with the scaling symmetry provides the following system of first
order ODEs%
\begin{eqnarray}
H_{t} &=&2H\left( \alpha U-V\right) ,  \label{se.05} \\
U_{t} &=&\alpha H^{2}+u\left( \alpha U-V\right) +fV,  \label{se.06} \\
V_{t} &=&-H^{2}-v\left( \alpha U-V\right) -fU,  \label{se.08}
\end{eqnarray}%
where $h=wH,~u=wU$ and $v=wU.~$The latter system is integrable and can be
solved with quadratures.

Reduction with respect the symmetry vector $\left( \alpha \sin \left(
ft\right) -\cos \left( ft\right) \right) \partial _{w}-f\left( \cos \left(
ft\right) \partial _{u}-\sin \left( ft\right) \partial _{v}\right) $ we find
the reduced system%
\begin{eqnarray}
\frac{H_{t}}{H} &=&-\frac{\alpha \cos \left( ft\right) +\sin \left(
ft\right) }{\cos \left( ft\right) -\alpha \sin \left( ft\right) },
\label{se.10} \\
U_{t} &=&-\alpha f\frac{\sin \left( ft\right) V-\cos \left( ft\right) U}{%
\cos \left( ft\right) -\alpha \sin \left( ft\right) },  \label{se.11} \\
V_{t} &=&-f\frac{\sin \left( ft\right) V-\cos \left( ft\right) U}{\cos
\left( ft\right) -\alpha \sin \left( ft\right) },  \label{se.12}
\end{eqnarray}%
where now
\begin{eqnarray}
h &=&H\left( t\right) ~,~ \\
u &=&\frac{\cos \left( ft\right) }{\cos \left( ft\right) -\alpha \sin \left(
ft\right) }fw+U\left( t\right) , \\
v &=&-\frac{\sin \left( ft\right) }{\cos \left( ft\right) -\alpha \sin
\left( ft\right) }fw+V\left( t\right) .
\end{eqnarray}

System (\ref{se.10}), (\ref{se.11}), (\ref{se.12}) is integrable and the
solution is expressed in terms of quadratures.

\section{Conclusions}

\label{sec5}

In this work, we determined the one-dimensional optimal system for the
two-dimensional ideal gas equations. The nonrotating system it was found
that is invariant under an eight-dimensional group of one-parameter point
transformations. and there are twenty-three independent one-dimensional Lie
algebras. One the other hand, when the Coriolis force is introduced, the
dynamical admits seven Lie point symmetries and twenty one-dimensional Lie
algebras.

For all the independent Lie algebras we determined all the invariant
functions which correspond to all the independent similarity solutions.

In a future work we plan to classify all the independent one-dimensional Lie
algebras which lead to analytic forms for the similarity solutions.

\textbf{Acknowledgements}

The author wants to thank Professor Kevin Duffy for gracious hospitality and
Professor PGL Leach for a fruitful discussion on the subject.

\appendix

\section{Tables}

\label{app1}

In this appendix we present the tables \ref{tabl2}, \ref{tabl2a}, \ref%
{tabl2b}, \ref{tabl4}, \ref{tabl4a} and \ref{tabl4b} which are refereed in
the main article.

\begin{landscape}
\begin{table}[tbp] \centering%
\caption{Adjoint representation of the admitted Lie point symmetries for the
nonrotating 2D Shallow water}%
\begin{tabular}{ccccccccc}
\hline\hline
$Ad\left( e^{\left( \varepsilon \mathbf{X}_{i}\right) }\right) \mathbf{X}%
_{j} $ & $\mathbf{X}_{1}$ & $\mathbf{X}_{2}$ & $\mathbf{X}_{3}$ & $\mathbf{X}%
_{4}$ & $\mathbf{X}_{5}$ & $\mathbf{X}_{6}$ & $\mathbf{X}_{7}$ & $\mathbf{X}%
_{8}$ \\
$\mathbf{X}_{1}$ & $X_{1}$ & $X_{2}$ & $X_{3}$ & $X_{4}-\varepsilon X_{2}$ &
$X_{5}-\varepsilon X_{3}$ & $X_{6}$ & $X_{7}+\varepsilon \left( \gamma
-1\right) X_{1}$ & $X_{8}$ \\
$\mathbf{X}_{2}$ & $X_{1}$ & $X_{2}$ & $X_{3}$ & $X_{4}$ & $X_{5}$ & $%
X_{6}+\varepsilon X_{3}$ & $X_{7}$ & $X_{8}-\varepsilon \left( \gamma
-1\right) X_{2}$ \\
$\mathbf{X}_{3}$ & $X_{1}$ & $X_{2}$ & $X_{3}$ & $X_{4}$ & $X_{5}$ & $%
X_{6}-\varepsilon X_{2}$ & $X_{7}$ & $X_{8}-\varepsilon \left( \gamma
-1\right) X_{3}$ \\
$\mathbf{X}_{4}$ & $X_{1}+\varepsilon X_{2}$ & $X_{2}$ & $X_{3}$ & $X_{4}$ &
$X_{5}$ & $X_{6}+\varepsilon X_{5}$ & $X_{7}-\varepsilon \left( \gamma
-1\right) X_{4}$ & $X_{8}-\varepsilon \left( \gamma -1\right) X_{4}$ \\
$\mathbf{X}_{5}$ & $X_{1}+\varepsilon X_{3}$ & $X_{2}$ & $X_{3}$ & $X_{4}$ &
$X_{5}$ & $X_{6}-\varepsilon X_{4}$ & $X_{7}-\varepsilon \left( \gamma
-1\right) X_{5}$ & $X_{8}-\varepsilon \left( \gamma -1\right) X_{5}$ \\
$\mathbf{X}_{6}$ & $X_{1}$ & $X_{2}\cos \varepsilon -X_{3}\sin \varepsilon $
& $X_{2}\sin \varepsilon +X_{3}\cos \varepsilon $ & $X_{4}\cos \varepsilon
-X_{5}\sin \varepsilon $ & $X_{4}\sin \varepsilon +X_{5}\cos \varepsilon $ &
$X_{6}$ & $X_{7}$ & $X_{8}$ \\
$\mathbf{X}_{7}$ & $e^{-\left( \gamma -1\right) \varepsilon }X_{1}$ & $X_{2}$
& $X_{3}$ & $e^{-\left( \gamma -1\right) \varepsilon }X_{4}$ & $e^{-\left(
\gamma -1\right) \varepsilon }X_{5}$ & $X_{6}$ & $X_{7}$ & $X_{8}$ \\
$\mathbf{X}_{8}$ & $X_{1}$ & $e^{\left( \gamma -1\right) \varepsilon }X_{3}$
& $e^{\left( \gamma -1\right) \varepsilon }X_{4}$ & $e^{-\left( \gamma
-1\right) \varepsilon }X_{4}$ & $e^{-\left( \gamma -1\right) \varepsilon
}X_{5}$ & $X_{6}$ & $X_{7}$ & $X_{8}$ \\ \hline\hline
\end{tabular}%
\label{tabl2}%
\end{table}
\end{landscape}%

\begin{landscape}
\begin{table}[tbp] \centering%
\caption{Lie invariants for the optimal system of the nonrotating system}%
\begin{tabular}{cc}
\hline\hline
\textbf{Symmetry} & \textbf{\ Invariants} \\
$\mathbf{X}_{1}$ & $x,~y,~h\left( x,y\right) ,~u\left( x,y\right) ,~v\left(
x,y\right) $ \\
$\mathbf{X}_{2}$ & $t,~y,~h\left( t,y\right) ,~u\left( t,y\right) ,~v\left(
t,y\right) $ \\
$\mathbf{X}_{3}$ & $t,~x,~h\left( t,x\right) ,~u\left( t,x\right) ,~v\left(
t,x\right) $ \\
$\mathbf{X}_{4}$ & $t,~y,~h\left( t,y\right) ,~\frac{x}{t}+U\left(
t,y\right) ,~v\left( t,y\right) $ \\
$\mathbf{X}_{5}$ & $t,~x,~h\left( t,x\right) ,~u\left( t,x\right) ,~\frac{y}{%
t}+V\left( t,x\right) $ \\
$\mathbf{X}_{6}$ & $t,~x^{2}+y^{2},~h\left( t,x^{2}+y^{2}\right) ,~\frac{%
xU\left( t,x^{2}+y^{2}\right) +yV\left( t,x^{2}+y^{2}\right) }{\sqrt{%
x^{2}+y^{2}}}~,~\frac{yU\left( t,x^{2}+y^{2}\right) -xV\left(
t,x^{2}+y^{2}\right) }{\sqrt{x^{2}+y^{2}}}$ \\
$\mathbf{X}_{7}$ & $\frac{x}{t},~\frac{y}{t},~h\left( \frac{x}{t},\frac{y}{t}%
\right) ,~u\left( \frac{x}{t},\frac{y}{t}\right) ,~v\left( \frac{x}{t},\frac{%
y}{t}\right) $ \\
$\mathbf{X}_{8}$ & $H\left( x,y\right) t^{\frac{2}{1-\gamma }},~U\left(
x,y\right) t^{-1}~,~V\left( x,y\right) t^{-1}$ \\
$\xi _{\left( 12\right) }$ & $x-\alpha t,~y,~h\left( x-\alpha t,y\right)
,~u\left( x-\alpha t,y\right) ,~v\left( x-\alpha t,y\right) $ \\
$\xi _{\left( 13\right) }$ & $x,~y-\alpha t,~h\left( x,y-\alpha t\right)
,~u\left( x,y-\alpha t\right) ,~v\left( x,y-\alpha t\right) $ \\
$\xi _{\left( 14\right) }$ & $x-\frac{\alpha }{2}t^{2},~y,~h\left( x-\frac{%
\alpha }{2}t^{2},y\right) ,~u\left( x-\frac{\alpha }{2}t^{2},y\right)
,~v\left( x-\frac{\alpha }{2}t^{2},y\right) $ \\
$\xi _{\left( 15\right) }$ & $x,~y-\frac{\alpha }{2}t^{2},~h\left( x,y-\frac{%
\alpha }{2}t^{2}\right) ,~u\left( x,y-\frac{\alpha }{2}t^{2}\right)
,~v\left( x,y-\frac{\alpha }{2}t^{2}\right) ~$ \\ \hline\hline
\end{tabular}%
\label{tabl2a}%
\end{table}
\end{landscape}%

\begin{landscape}
\begin{table}[tbp] \centering%
\caption{Lie invariants for the optimal system of the nonrotating system}%
\begin{tabular}{cc}
\hline\hline
\textbf{Symmetry} & \textbf{\ Invariants} \\
$\xi _{\left( 16\right) }$ & $~~%
\begin{tabular}{l}
$t,~e^{-\alpha t}\left( x^{2}+y^{2}\right) ,~u\left( t,e^{-\alpha
t}x^{2}+y^{2}\right) \cos \left( \alpha t\right) +v\left( t,e^{-\alpha
t}x^{2}+y^{2}\right) \sin \left( \alpha t\right) $ \\
$\frac{y}{x},~h\left( e^{-\alpha t}x^{2}+y^{2},\frac{y}{x}\right) ,~u\left(
t,e^{-\alpha t}x^{2}+y^{2}\right) \sin \left( \alpha t\right) -v\left(
t,e^{-\alpha t}x^{2}+y^{2}\right) \cos \left( \alpha t\right) $%
\end{tabular}%
$ \\
$\xi _{\left( 23\right) }$ & $t,~x-\alpha y,~h\left( t,x-\alpha y\right)
,~u\left( t,x-\alpha y\right) ,v\left( t,x-\alpha y\right) $ \\
$\xi _{\left( 34\right) }$ & $t,~y-\frac{x}{\alpha t},~h\left( t,y-\frac{x}{%
\alpha t}\right) ,~u\left( t,y-\frac{x}{\alpha t}\right) ,~v\left( t,y-\frac{%
x}{\alpha t}\right) $ \\
$\xi _{\left( 25\right) }$ & $t,~y-\alpha tx,~h\left( t,y-\alpha tx\right)
,~u\left( t,y-\alpha tx\right) ,~v\left( t,y-\alpha tx\right) $ \\
$\xi _{\left( 45\right) }$ & $t,~y-\alpha x,~h\left( t,y-\alpha x\right)
,~\alpha \frac{x}{t}+U\left( t,y-\alpha x\right) ,~\alpha \frac{x}{t}%
+V\left( t,y-\alpha x\right) $ \\
$\xi _{\left( 123\right) }$ & $t-\alpha x,~t-\beta y,~h\left( t-\alpha
x,t-\beta y\right) ,~u\left( t-\alpha x,t-\beta y\right) ,v\left( t-\alpha
x,t-\beta y\right) $ \\
$\xi _{\left( 145\right) }$ & $x-\frac{\alpha }{2}t^{2},~y-\frac{\beta }{2}%
t^{2},~h\left( x-\frac{\alpha }{2}t^{2},y-\frac{\beta }{2}t^{2}\right)
,~\alpha t+U\left( x-\frac{\alpha }{2}t^{2},~y-\frac{\beta }{2}t^{2}\right)
,~\beta t+V\left( x-\frac{\alpha }{2}t^{2},~y-\frac{\beta }{2}t^{2}\right) $
\\
$\xi _{\left( 125\right) }$ & \thinspace $x-\alpha t,~y-\frac{\beta }{2}%
t^{2},~h\left( x-\alpha t,y-\frac{\beta }{2}t^{2}\right) ,~u\left( x-\alpha
t,y-\frac{\beta }{2}t^{2}\right) ,~\beta t+V\left( x-\alpha t,y-\frac{\beta
}{2}t^{2}\right) $ \\
$\xi _{\left( 134\right) }$ & $x-\frac{\beta }{2}t^{2},~y-\alpha t,~h\left(
x-\frac{\beta }{2}t^{2},y-\alpha t\right) ,~\beta t+U\left( x-\frac{\beta }{2%
}t^{2},y-\alpha t\right) ,~V\left( x-\frac{\beta }{2}t^{2},y-\alpha t\right)
$ \\
$\xi _{\left( 67\right) }$ &
\begin{tabular}{l}
$\frac{\ln t}{\alpha },w=~\frac{t^{-\frac{\alpha +\sqrt{\alpha \left( \alpha
-4\right) -4}}{2\alpha }}}{2\sqrt{\alpha \left( \alpha -4\right) -4}}\left(
x-\left( \alpha +\sqrt{\alpha \left( \alpha -4\right) -4}\right) y\right)
,~z=~\frac{t^{-\frac{\alpha +\sqrt{\alpha \left( \alpha -4\right) -4}}{%
2\alpha }}}{2\sqrt{\alpha \left( \alpha -4\right) -4}}\left( x+\left( \alpha
+\sqrt{\alpha \left( \alpha -4\right) -4}\right) y\right) $ \\
$~h\left( w,z\right) ,~U\left( w,z\right) \sin \left( \frac{\ln t}{\alpha }%
\right) +V\left( w,z\right) \sin \left( \frac{\ln t}{\alpha }\right)
~,~U\left( w,z\right) \cos \left( \frac{\ln t}{\alpha }\right) -~V\left(
w,z\right) \sin \left( \frac{\ln t}{\alpha }\right) $%
\end{tabular}%
$~$ \\
$\xi _{\left( 68\right) }$ & $t,~x^{2}+y^{2}~,~x^{-\frac{2}{\gamma -1}%
}h\left( t,x^{2}+y^{2}\right) ~,~\frac{U\left( t,x^{2}+y^{2}\right) \cos
\left( \frac{\ln x}{\alpha }\right) +V\left( t,x^{2}+y^{2}\right) \sin
\left( \frac{\ln x}{\alpha }\right) }{x}~,~~\frac{U\left(
t,x^{2}+y^{2}\right) \sin \left( \frac{\ln x}{\alpha }\right) -V\left(
t,x^{2}+y^{2}\right) \cos \left( \frac{\ln x}{\alpha }\right) }{x}$ \\
$\xi _{\left( 78\right) }$ & $w=xt^{-\frac{\left( \gamma -1\right) }{\alpha
\left( \gamma -1\right) -2}},~z=yt^{-\frac{\left( \gamma -1\right) }{\alpha
\left( \gamma -1\right) -2}},~t^{-\frac{2\alpha }{\alpha \left( \gamma
-1\right) -2}}h\left( w,z\right) ,t^{-\frac{\left( \gamma -1\right) \alpha }{%
\alpha \left( \gamma -1\right) -2}}u\left( w,z\right) ,~t^{-\frac{\left(
\gamma -1\right) \alpha }{\alpha \left( \gamma -1\right) -2}}v\left(
w,z\right) $ \\
$\xi _{\left( 678\right) }$ &
\begin{tabular}{l}
$t~,~t^{-1-\beta }\left( x^{2}+y^{2}\right) ~,t^{-\beta }\left( U\left(
t,x^{2}+y^{2}\right) \sin \left( \alpha t\right) +V\left(
t,x^{2}+y^{2}\right) \cos \left( \alpha t\right) \right) $ \\
$~t^{-\frac{2\beta }{\gamma -1}}H\left( t,x^{2}+y^{2}\right) ~,~t^{-\beta
}\left( U\left( t,x^{2}+y^{2}\right) \cos \left( \alpha t\right) -V\left(
t,x^{2}+y^{2}\right) \sin \left( \alpha t\right) \right) $%
\end{tabular}%
$~$ \\ \hline\hline
\end{tabular}%
\label{tabl2b}%
\end{table}
\end{landscape}%

\bigskip

\begin{landscape}
\begin{table}[tbp] \centering%
\caption{Adjoint representation of the admitted Lie point symmetries for the
rotating 2D Shallow water}%
\begin{tabular}{cccccccc}
\hline\hline
$Ad\left( e^{\left( \varepsilon \mathbf{Y}_{i}\right) }\right) \mathbf{Y}%
_{j} $ & $\mathbf{Y}_{1}$ & $\mathbf{Y}_{2}$ & $\mathbf{Y}_{3}$ & $\mathbf{Y}%
_{4}$ & $\mathbf{Y}_{5}$ & $\mathbf{Y}_{6}$ & $\mathbf{Y}_{7}$ \\
$\mathbf{Y}_{1}$ & $Y_{1}$ & $Y_{2}$ & $Y_{3}$ & $Y_{4}$ & $Y_{5}\cos \left(
f\varepsilon \right) -Y_{6}\sin \left( f\varepsilon \right) $ & $Y_{5}\sin
\left( f\varepsilon \right) +Y_{6}\cos \left( f\varepsilon \right) $ & $%
Y_{7} $ \\
$\mathbf{Y}_{2}$ & $Y_{1}$ & $Y_{2}$ & $Y_{3}$ & $Y_{4}+\varepsilon Y_{3}$ &
$Y_{5}$ & $Y_{6}$ & $Y_{7}-\varepsilon \left( \gamma -1\right) Y_{2}$ \\
$\mathbf{Y}_{3}$ & $Y_{1}$ & $Y_{2}$ & $Y_{3}$ & $Y_{4}-\varepsilon Y_{2}$ &
$Y_{5}$ & $Y_{6}$ & $Y_{7}-\varepsilon \left( \gamma -1\right) Y_{3}$ \\
$\mathbf{Y}_{4}$ & $Y_{1}$ & $Y_{2}\cos \varepsilon -Y_{3}\sin \varepsilon $
& $Y_{2}\sin \varepsilon +Y_{3}\cos \varepsilon $ & $Y_{4}$ & $Y_{5}\cos
\varepsilon +Y_{6}\sin \varepsilon $ & $Y_{6}\cos \varepsilon -Y_{5}\sin
\varepsilon $ & $Y_{7}$ \\
$\mathbf{Y}_{5}$ & $Y_{1}+f\varepsilon Y_{6}$ & $Y_{2}$ & $Y_{3}$ & $%
Y_{4}-\varepsilon Y_{6}$ & $Y_{5}$ & $Y_{6}$ & $Y_{7}-\varepsilon \left(
\gamma -1\right) Y_{5}$ \\
$\mathbf{Y}_{6}$ & $Y_{1}-f\varepsilon Y_{5}$ & $Y_{2}$ & $Y_{3}$ & $%
Y_{4}+\varepsilon Y_{5}$ & $Y_{5}$ & $Y_{6}$ & $Y_{7}-\varepsilon \left(
\gamma -1\right) Y_{6}$ \\
$\mathbf{Y}_{7}$ & $Y_{1}$ & $e^{\left( \gamma -1\right) \varepsilon }Y_{2}$
& $e^{\left( \gamma -1\right) \varepsilon }Y_{3}$ & $Y_{4}$ & $e^{\left(
\gamma -1\right) \varepsilon }Y_{5}$ & $e^{\left( \gamma -1\right)
\varepsilon }Y_{6}$ & $Y_{7}$ \\ \hline\hline
\end{tabular}%
\label{tabl4}%
\end{table}
\end{landscape}%

\begin{landscape}
\begin{table}[tbp] \centering%
\caption{Lie invariants for the optimal system of the rotating system}%
\begin{tabular}{cc}
\hline\hline
\textbf{Symmetry} & \textbf{\ Invariants} \\
$\mathbf{Y}_{1}$ & $x,~y,~h\left( x,y\right) ,~u\left( x,y\right) ,~v\left(
x,y\right) $ \\
$\mathbf{Y}_{2}$ & $t,~y,~h\left( t,y\right) ,~u\left( t,y\right) ,~v\left(
t,y\right) $ \\
$\mathbf{Y}_{3}$ & $t,~x,~h\left( t,x\right) ,~u\left( t,x\right) ,~v\left(
t,x\right) $ \\
$\mathbf{Y}_{4}$ & $t,~x^{2}+y^{2},~h\left( t,x^{2}+y^{2}\right) ,~\frac{%
xU\left( t,x^{2}+y^{2}\right) +yV\left( t,x^{2}+y^{2}\right) }{\sqrt{%
x^{2}+y^{2}}}~,~\frac{yU\left( t,x^{2}+y^{2}\right) -xV\left(
t,x^{2}+y^{2}\right) }{\sqrt{x^{2}+y^{2}}}$ \\
$\mathbf{Y}_{5}$ & $t,~x\cot \left( ft\right) -y~,~h\left( t,x\cot \left(
ft\right) -y\right) ,~fx\cot \left( ft\right) +U\left( t,x\cot \left(
ft\right) -y\right) ,~-fx+V\left( t,x\cot \left( ft\right) -y\right) $ \\
$\mathbf{Y}_{6}$ & $t,~x\tan \left( ft\right) +y,~h\left( t,x\tan \left(
ft\right) +y\right) ,~-fx\tan \left( ft\right) +U\left( t,x\tan \left(
ft\right) +y\right) ,~-fx+V\left( t,x\tan \left( ft\right) +y\right) $ \\
$\mathbf{Y}_{7}$ & $\frac{x}{t},~\frac{y}{t},~h\left( \frac{x}{t},\frac{y}{t}%
\right) ,~u\left( \frac{x}{t},\frac{y}{t}\right) ,~v\left( \frac{x}{t},\frac{%
y}{t}\right) $ \\
$\chi _{\left( 12\right) }$ & $x-\alpha t,~y,~h\left( x-\alpha t,y\right)
,~u\left( x-\alpha t,y\right) ,~v\left( x-\alpha t,y\right) $ \\
$\chi _{\left( 13\right) }$ & $x,~y-\alpha t,~h\left( x,y-\alpha t\right)
,~u\left( x,y-\alpha t\right) ,~v\left( x,y-\alpha t\right) $ \\
$\chi _{\left( 14\right) }$ & $~~%
\begin{tabular}{l}
$t,~e^{-\alpha t}\left( x^{2}+y^{2}\right) ,~u\left( t,e^{-\alpha
t}x^{2}+y^{2}\right) \cos \left( \alpha t\right) +v\left( t,e^{-\alpha
t}x^{2}+y^{2}\right) \sin \left( \alpha t\right) $ \\
$\frac{y}{x},~h\left( e^{-\alpha t}x^{2}+y^{2},\frac{y}{x}\right) ,~u\left(
t,e^{-\alpha t}x^{2}+y^{2}\right) \sin \left( \alpha t\right) -v\left(
t,e^{-\alpha t}x^{2}+y^{2}\right) \cos \left( \alpha t\right) $%
\end{tabular}%
$ \\ \hline\hline
\end{tabular}%
\label{tabl4a}%
\end{table}
\end{landscape}%

\begin{landscape}
\begin{table}[tbp] \centering%
\caption{Lie invariants for the optimal system of the rotating system}%
\begin{tabular}{cc}
\hline\hline
\textbf{Symmetry} & \textbf{\ Invariants} \\
$\chi _{\left( 15\right) }$ &
\begin{tabular}{l}
$x+\frac{\alpha }{f}\cos \left( ft\right) ,~y-\frac{\alpha }{f}\sin \left(
ft\right) ~,~h\left( \,x+\frac{\alpha }{f}\cos \left( ft\right) ,y-\frac{%
\alpha }{f}\sin \left( ft\right) \right) ,~$ \\
$\alpha \sin \left( ft\right) +U\left( \,x+\frac{\alpha }{f}\cos \left(
ft\right) ,y-\frac{\alpha }{f}\sin \left( ft\right) \right) ,~\alpha \cos
\left( ft\right) +V\left( \,x+\frac{\alpha }{f}\cos \left( ft\right) ,y-%
\frac{\alpha }{f}\sin \left( ft\right) \right) $%
\end{tabular}%
\thinspace \\
$\chi _{\left( 16\right) }$ &
\begin{tabular}{l}
$x-\frac{\alpha }{f}\sin \left( ft\right) ,~y-\frac{\alpha }{f}\cos \left(
ft\right) ~,~h\left( x-\frac{\alpha }{f}\sin \left( ft\right) ,y-\frac{%
\alpha }{f}\cos \left( ft\right) \right) ,~$ \\
$\alpha \cos \left( ft\right) +U\left( x-\frac{\alpha }{f}\sin \left(
ft\right) ,y-\frac{\alpha }{f}\cos \left( ft\right) \right) ,~-\alpha \sin
\left( ft\right) +V\left( x-\frac{\alpha }{f}\sin \left( ft\right) ,y-\frac{%
\alpha }{f}\cos \left( ft\right) \right) $%
\end{tabular}
\\
$\chi _{\left( 17\right) }$ & $xe^{-\alpha t},~ye^{-\alpha t},~e^{\frac{%
2\alpha }{\gamma -1}t}h\left( xe^{-\alpha t},ye^{-\alpha t}\right)
,~e^{\alpha t}u\left( xe^{-\alpha t},ye^{-\alpha t}\right) ,~e^{\alpha
t}v\left( xe^{-\alpha t},ye^{-\alpha t}\right) $ \\
$\chi _{\left( 23\right) }$ & $t,~x-\alpha y,~h\left( t,x-\alpha y\right)
,~u\left( t,x-\alpha y\right) ,~v\left( t,x-\alpha y\right) $ \\
$\chi _{\left( 45\right) }$ & $t,~w=\left( x^{2}+y^{2}-2x~\cos \left(
ft\right) +2y\sin \left( ft\right) \right) ,~\frac{U\left( t,w\right) +f\sin
\left( ft\right) }{V\left( t,w\right) +f\cos \left( ft\right) },~\frac{%
U\left( t,w\right) ^{2}+V\left( t,w\right) ^{2}}{2}+f\left( U\left(
t,w\right) \sin \left( ft\right) +V\left( t,w\right) \cos \left( ft\right)
\right) $ \\
$\chi _{\left( 46\right) }$ & $t,~w=\left( x^{2}+y^{2}-2x~\sin \left(
ft\right) -2y\cos \left( ft\right) \right) ,~\frac{U\left( t,w\right) -f\cos
\left( ft\right) }{V\left( t,w\right) +f\sin \left( ft\right) },~\frac{%
U\left( t,w\right) ^{2}+V\left( t,w\right) ^{2}}{2}+f\left( V\left(
t,w\right) \sin \left( ft\right) -U\left( t,w\right) \cos \left( ft\right)
\right) $ \\
$\chi _{\left( 56\right) }$ & $t,~z=y-\frac{x\left( \cos \left( ft\right)
-\alpha \sin \left( ft\right) \right) }{\sin \left( ft\right) +\alpha \cos
\left( ft\right) },~h\left( t,z\right) ,~f\frac{x\left( \cos \left(
ft\right) -\alpha \sin \left( ft\right) \right) }{\sin \left( ft\right)
+\alpha \cos \left( ft\right) }+U\left( t,z\right) ,~-x+V\left( t,z\right) $
\\
$\chi _{\left( 47\right) }$ & $t,~x^{2}+y^{2}~,~x^{-\frac{2}{\gamma -1}%
}h\left( t,x^{2}+y^{2}\right) ~,~\frac{U\left( t,x^{2}+y^{2}\right) \cos
\left( \frac{\ln x}{\alpha }\right) +V\left( t,x^{2}+y^{2}\right) \sin
\left( \frac{\ln x}{\alpha }\right) }{x}~,~~\frac{U\left(
t,x^{2}+y^{2}\right) \sin \left( \frac{\ln x}{\alpha }\right) -V\left(
t,x^{2}+y^{2}\right) \cos \left( \frac{\ln x}{\alpha }\right) }{x}$ \\
$\chi _{\left( 123\right) }$ & $t-\alpha x,~t-\beta y,~h\left( t-\alpha
x,t-\beta y\right) ,~u\left( t-\alpha x,t-\beta y\right) ,v\left( t-\alpha
x,t-\beta y\right) $ \\
$\chi _{\left( 147\right) }$ & $\,%
\begin{tabular}{l}
$z=e^{-t\left( \gamma -1\right) }\left( x\cos t-y\sin t\right)
~,~w=e^{-t\left( \gamma -1\right) }\left( y\cos t+x\sin t\right) ~,$ \\
$e^{-t\left( \gamma -1\right) }h\left( z,w\right) ,~e^{-t\left( \gamma
-1\right) }\left( U\left( z,w\right) \cos t-V\left( z,w\right) \sin t\right)
~,~e^{-t\left( \gamma -1\right) }\left( U\left( z,w\right) \sin t+V\left(
z,w\right) \cos t\right) $%
\end{tabular}%
~$ \\ \hline\hline
\end{tabular}%
\label{tabl4b}%
\end{table}
\end{landscape}%

\end{document}